\documentstyle[12pt]{article}
\newcommand{\ep}{\varepsilon}
\newcommand{\Proj}{{\rm Proj}\ }
\newcommand{\mod}{{\rm mod}}
\newcommand{\Dif}{{\rm Dif}}
\newcommand{\Differ}{{\rm Differ}}
\newcommand{\Difference}{{\rm Difference}}
\newcommand{\Change}{{\rm Change}}
\renewcommand{\Re}{{\rm Re}}
\newcommand{\Res}{{\rm Res}}
\newcommand{\Sig}{{\rm Sig}}
\newcommand{\U}{{\rm U}}
\newcommand{\Check}{{\rm Check}}
\newcommand{\Anc}{{\rm Anc}}
\renewcommand{\d}{\delta}

\newcommand{\App}{{\rm App}}
\newcommand{\Spectr}{{\rm Spectr}}
\newcommand{\State}{{\rm State}}
\newcommand{\SignDif}{{\rm SignDif}}
\newcommand{\GenFreq}{{\rm GenFreq}}
\newcommand{\GenTimeFreq}{{\rm GenTimeFreq}}
\newcommand{\Dist}{{\rm Dist}}
\newcommand{\Turn}{{\rm Turn}}
\newcommand{\Inv}{{\rm Inv}}
\newcommand{\Si}{{\rm Si}}
\newcommand{\GenArg}{{\rm GenArg}}
\newcommand{\GenTimeArg}{{\rm GenTimeArg}}

\newcommand{\Conc}{{\rm Conc}}
\newcommand{\Sign}{{\rm Sign}}
\newcommand{\GenTime}{{\rm GenTime}}
\newcommand{\QFT}{{\rm QFT}}
\newcommand{\Enh}{{\rm Enh}}
\newcommand{\Gen}{{\rm Gen}}
\newcommand{\Rev}{{\rm Rev}}
\newcommand{\Rest}{{\rm Rest}}

\newcommand{\card}{{\rm card}}

\newcommand{\e}{\epsilon}

\newcommand{\St}{{\rm St}}

\newcommand{\SignGoodFreq}{{\rm SignGoodFreq}}

\newcommand{\ar}{\longrightarrow}
\newcommand{\n}{\smallskip}
\newcommand{\nn}{\medskip}

\newcommand{\Rec}{{\rm Rec}}
\newcommand{\w}{\omega}

\newcommand{\la}{\lambda}
\renewcommand{\a}{\alpha}

\title{Quantum recognition of eigenvalues, structure of devices and 
thermodynamic properties}

\author{ Yuri I. Ozhigov\thanks{Institute of Physics
and Technology, Moscow, email: ozhigov@ftian.oivta.ru}}

\begin{document}

\maketitle

\begin{abstract}

Quantum algorithms speeding up classical counterparts are proposed for the 
problems:

1. Recognition of eigenvalues with fixed precision. Given a quantum circuit 
generating unitary mapping $U$ and a complex number the problem is 
to determine is it an eigenvalue of $U$ or not. 

2. Given a molecular structure find 
thermodynamic functions like partitioning function, entropy, etc. for 
a gas consisting of such molecules.
 
3. Recognition of molecular structures. Find a molecular structure 
given its spectrum.

4. Recognition of electronic devices. Given an electronic device that can be 
used only as a black box how to recognize its internal construction? 

We consider mainly structures generating sparse spectrums.
These algorithms require the time from about square root to logarithm of the 
time of classical analogs and for the first three problems give exponential 
memory saving. Say, the time required for distinguishing two 
devices with the same given spectrum is about seventh root of the time of direct  classical method, for the 
recognition of eigenvalue - about sixth root. Thus microscopic quantum devices 
can recognize molecular structures and physical properties of environment 
faster than big classical computers. 

\end{abstract}

{\bf Contents}
\begin{enumerate}
\item Electronic devices and quantum computations
\begin{enumerate}
\item Statements of problems and outline of the work
\item Abstract model of QC. "Plug and play" technology
\end{enumerate}
\item Two basic quantum tricks
\begin{enumerate}
\item GSA and amplitude amplification
\item Revealing of eigenvalues
\end{enumerate}
\item Problems of recognition
\begin{enumerate}
\item Getting eigenvectors and recognition of eigenvalues
\item Finding of thermodynamic functions
\item Recognition of molecular structures
\item Distinguishing of eigenvectors of two operators with the same 
eigenvalue
\item Recognition of electronic devices circuits
\item Advantages of algorithms over classical counterparts
\end{enumerate}
\item Conclusion
\item Acknowledgments
\item References
\end{enumerate}

\section{Electronic devices and quantum computations}

\subsection{Statements of problems and outline of the work}

The aim of this paper is to build effective quantum algorithms for the 
problems of the following types:

\begin{itemize}
\item Given a quantum gate array generating unitary operator $U$ and a complex 
number $\w$ how to determine is it an eigenvalue of $U$ or not (precision of
 determining eigenvalues is fixed)?

\item How to recognize a structure of unknown electronic or molecular device given
 only access to its function? 
\end{itemize}

Here the first problem will be an important intermediate step in the solution
 of the second \footnote{A straightforward calculation shows
that the simulation of evolution generated by a given Hamiltonian up to a time 
instant $\tau$ with a 
fixed accuracy requires of order $\tau^2$ steps on a quantum computer. This 
means that all results
 of the paper can be generalized to arbitrary quantum systems. This subject 
will be elaborated in more details in another paper.}. Consider them sequentially. 

{\bf Recognition of eigenvalues.} This problem is closely connected with 
finding of eigenvalues distribution or density of states (DOS) that is energy 
levels $E_0 <E_1 <\ldots$ and dimensions of the corresponding subspaces
$d_0 , d_1 , \ldots$. DOS plays a key role in
 calculation of thermodynamic functions given by 
\begin{equation}
F=\sum\limits_j a(j)d_j e^{-\frac{E_j
}{k_B T}}
\label{Therm}
\end{equation}
for some values $a(j)$ so that the summands quickly converge to zero. Say, if 
all $a(j)=1$ this expression gives the partition function $Q$, 
if $a(j)=E_j /Q$ it gives an average energy, if $a(j)=-\frac{k_B }{Q} 
\ln (e^{-E_j /k_B T}/Q)$ - an entropy. Having an efficient method of 
finding $d_j$ we would be able to obtain thermodynamic functions and 
to determine important properties of environment consisting of such 
molecules like heat capacity. The best known classical method of finding 
DOS was proposed by Hams and Raedt in the work \cite{HR}. Their method 
requires the time of order dimension $N$ of the space of states and the 
memory of the same order (whereas the direct method of calculation eigenvalues 
requires the time of order $N^3$). The first quantum algorithm for this problem 
proposed by 
Abrams and Lloyd in \cite{AL} requires the same time $O(N)$ and logarithmic memory. 
The method proposed in the present work requires the time of order square 
root of classical and memory of order $\ln^2 N$. 

The idea of our approach is the following. We shall use combination of Grover 
search algorithm (GSA), revealing eigenvalues by Abrams and Lloyd method (\cite{AL})
 and universal quantum function of application $\App$. Abrams and Lloyd method of revealing eigenvalues is based on the application of $U$ controlled by ancillary qubit $\a$ as
$$
U_{cond} |x,\a \rangle \ar\left\{
\begin{array}{cc}
|U \ x,\a\rangle,\ &\mbox{if} \ \a =1,\\
|x,\a\rangle\ &\mbox
{if} \ \a=0.
\end{array}
\right.
$$
Note that it is the direct generalization of Shor's trick which can be obtained 
if $U$ is a multiplication by a given integer modulo $q$ (\cite{Sh}).

{\bf Recognition of devices structure.} 

We shall 
tell apart two versions of this general problem: the recognition of molecular
structures and the recognition of electronic circuits. 

If we want to determine a molecular structure then it is natural to assume that its 
functionality is given as the spectrum of its Hamiltonian, e.g. the set of its energy levels. Thus 
here it
is required to find a quantum system whose Hamiltonian has a given 
spectrum. 

The problem of recognition of electronic circuits is stated otherwise. An
electronic device is thought of as a source of electromagnetic fields which
 can control some quantum system $Q$. Let such a field induces evolution of 
the system with Hamiltonian $H$ in time frame $\d t$. Thus we have a
correspondence: electronic device $\ar $ Hamiltonian, $\d t$. Evolution of a
quantum system $Q$ induced by this Hamiltonian can be presented as a
 unitary transform $U=e^{-\frac{i}{h} H\d t}$. Then, given a device $C$ and a value of time $t$ 
we can associate with it some unitary transformation $U_C$. Assume that
we have recognized a circuit $C$ if we find some circuit $C_1$ 
 such that $U_C =U_{C_1}$ with high accuracy. We shall write $U$ instead of
$U_C$ for the circuit 
$C$ that we want to recognize. But in fact we shall solve the more general 
problem when a tested device $C$ can be used as a black box acting on $n$ 
qubits as a function $U_C$ so that if $x$ is an input then $U_C |x\rangle$ 
is a
result of its action on this input. Here a tested device can contain 
its own quantum memory and it can be entangled with $Q$ in course of
fulfilling $U$ but this entanglement must be then eliminated. The existence of
 such entanglement means that this case cannot be described by the Hamiltonian
 of system $Q$. 
For the simplicity we assume that an 
unknown circuit is built of 
elementary functional elements from some fixed set $\{ E_1 , E_2 , \ldots ,
E_o \}$. The next natural
 assumption is that a size of circuit is limited by some constant $c$ so that 
our circuit is some unknown combination of $c$ functional elements. Denote by
$\cal E$ all circuits of the length $c$. We can encode such $C\in\cal E$ by a
 string $[C]$ of ones and zeroes so that decoding procedure is easy as well 
and we can immediately recreate a circuit given its code. Thus we can look
through all circuits looking through its codes. The same coding can be built 
for electronic devices. 

A straightforward solution of the problems is clear. For the problem of 
recognition of molecular structures all that we need is to be able to recognize
eigenvalues of transformation generated by a
given circuit. Each eigenvalue of unitary operator has the form
 $e^{2\pi i\w}$ where $\w$ is a real number from $[0,1)$ called frequency. 
In what follows by spectrum we mean a set of all frequencies. Let all 
frequencies are grouped near points of the form
$\frac{l}{M}$ where $M$ is not very big, $l=0,1,\ldots , M-1$. Assume that the acceptable precision of recognition of frequencies is $1/M$. 
Then having an algorithm for eigenvalue recognition we can apply it again and 
again constructing spectrums generated by all possible circuits and thus find a 
wanted circuit with given spectrum. If we need to recognize a circuit of electronic device we can
examine all possible circuits
taken in some order. Examination of one circuit means that we 
run it on all possible inputs one after another and compare the results with 
the corresponding result of a tested device action. 

For the problem of recognition of molecular structures our method requires the
 time of order sixth root of the time of direct classical method whereas memory 
saving is exponential. For the problem of recognition of electronic circuits our 
method gives 
at least square root time saving in the case when classical counterparts exist
 (this is the narrow formulation when a tested device generates classical 
mapping). But in general case an advantage may be more. For example we can tell apart two devices with the same spectrum in the time about seventh root of the time of naive bruit force.  

To recognize devices quantumly we must be able to store and fulfill operations 
on the codes of different circuits. This possibility is based
on the existence of a quantum analog of the universal Klini
 function. This is a unitary operator $\App$ such that for all quantum
devices $C$ and all inputs $x$ $\App |x,[C]\rangle =
|U_C \ x,[C]\rangle$. We assume that for the wide range of quantum devices $C$
with $c$ particles $C$ may be encoded as integer $[C]$ 
in time $O(c)$ so that the quantum complexity of $\App$ is $O(c)$ as well.

We shall consider here a particular case of the problem when all eigenvalues
of $U$ are known a-priori or can be obtained beforehand. This restriction is
not yet very constraining. To illustrate what kind of tasks we shall be able 
to solve by the proposed method consider a few examples of the problem 
of recognition of an electronic device whose spectrum is 
known. 

Recognition of quantum algorithms
designed as subroutines. Such algorithms must restore an input if we apply 
it twice. Computing a function
 $f$ they act as $|x,b\rangle \ar |x,b+ f(x)\ \mod\ 2\rangle$. All known
quantum algorithms can be presented in such form. For such quantum
algorithms their unitary transformation $U$ has only two eigenvalues: $1$
and $-1$. Given a controlling device for such algorithm (it may include 
classical elements and ancillary qubits as well) we can quickly recognize 
its construction. Alternatively, we can quickly find quantum or classical
algorithm for a given task. 

Consider a "classical" particular case of the recognition problem when $U$ 
maps
 each basic state to a basic state which means that the matrix of $U$ consists
of ones and zeroes and in addition $U$ equals $U^{-1}$. Here evident strategy 
of recognition takes of order $\card ({\cal E}) 2^n$ steps. This case of the
problem may be reformulated as the finding of such $t$ that for all $s$ some
given predicate $A(t,s)$ is true. This is the problem of verification of logical 
formulas. Quantum solution of it in a time about square root of classical time 
based plainly on Grover's trick was proposed in \cite{BCW}. This
 method doesn't work in the general case where $U_C$ is arbitrary 
involutive unitary transform, e.g. such that $U=U^{-1}$. Just this
general case is the subject of this work. Here we cannot recognize a 
circuit so easy as in "classical" case because
it is difficult to compare two quantum states $U_C  |x\rangle$ and $U  
|x\rangle$.

The general idea of our approach to the recognition of arbitrary electronic 
devices is the following. We shall include a device $C$ which structure we 
want to recognize into a classical controlling part of quantum computer.
 Thus a tested device generates unitary transformation on $n$ qubit system. 
Then reveal eigenvectors of $U$ using $U_{cond}$ by the method mentioned above and compare them with
eigenvectors of circuits from ${\cal E}$ choosing a circuit giving the best
approximation. 
Here GSA will be used in the last step and in the several intermediate steps.

{\bf Assumption about sparse spectrum}

In this paper we shall consider mainly circuits generating sparse spectrums. It means that spectrums 
of operators $U_C$ are so designed that the frequencies are grouped into 
groups such that a minimal distance between frequencies from the different 
groups is more than $1/M$ and a maximal distance between frequencies from the
same group is less than $1/L$. For the problems of eigenvalues and molecular 
structures recognition we require that $L=16M$ that is not yet very restricting. For the recognition of electronic devices we shall suppose $L\gg M$ that is more strong limitation. Spectrums are called sparse if $M=const$ when $N\ar\infty$. For sparse spectrums our algorithms show the best performance.

Spectrums that are not sparse are called dense. For dense spectrums our methods give the less advantage over classical algorithms (look at section 3.6). An example of dense spectrum: $\w_k =\frac{k}{N} ,\ k=0,1,\ldots ,N-1$. The similar problems for dense spectrums will be studied in one of the following papers.

 We write $\w' 
\approx\w$ iff $\w'$ and $\w$ belong to the same group. For the simplicity 
assume also that for each group of frequencies there exists a number of the
form $l/M$ disposed between some two frequencies of this group where $l$ is 
an integer less than $M$.

\subsection{An abstract model of QC. "Plug and play" technology}

To build algorithms recognizing circuits we need an abstract model of 
quantum computer (QC). QC consists of two parts: quantum and classical.
 Classical part exactly determines what unitary transformation must be
fulfilled at each time instant with quantum part and thus plays a role of 
controller for it. These unitary transformations are of two sorts: working 
transformations - which our computer performs itself and query transformations - 
induced by a tested device: 
$U$
or $U_{cond}$.

We can suppose that a quantum part $Q$ consists of nuclear spins or 
interacting dipoles (or some other quantum two levels systems) and a
classical part is a source of electromagnetic fields determining evolution 
of a quantum part.  The general 
form of a state of quantum part will be $\chi =\sum\limits_{i=0}^{2^{\nu}-1} 
\la_i e_i$ where the basic states of it $e_0 ,\ldots , e_{2^{\nu} -1}$ are
simply
strings of ones and zeroes of the length $\nu$ where $\nu > n$ is the size 
of
 quantum part which can contain some auxiliary qubits behind input for $U$, 
$\sum\limits_{i=0}^{2^{\nu}-1} |\la_i |^2 =1$, $N=2^n$ is the number of 
all classical input words for $U$.

Classical part determines when to "switch on" a tested device (usually it happens many times) and when to observe a result of computation. Observation of a 
state
$\chi$ gives every basic state $e_i$ with the corresponding probability
$|\la_i |^2$.

The problem of recognition of electronic devices presumes the so-called "plug 
and play" technology where a tested device is applied
only as a black box. 
If query transformations are only $U$ then our model evidently satisfies 
requirements of "plug and play" technology where we classically control when 
to switch on a tested device. An 
implementation of $U_{cond}$ in the framework of this technology is not so 
easy because it requires a quantum control on applications of the device \footnote{This would be evidently possible provided we have access to the 
internal details of our device and can quantumly control their work 
simultaneously. But this assumption contradicts to "plug and play" technology.}. 
Nevertheless
it is possible to implement $U_{cond}$ in the framework of "plug and play"
technology. This possibility will be substantiated in the following papers. 
Now we shall simply presume that it is possible. Such difficulty does not exist 
for the problems of eigenvalue and molecular structures recognition. Here we can 
manage without oracles at all because having an explicit form of a quantum gate 
array realizing a universal function of application $\App$ we can quantumly 
control its actions in each element separately and simultaneously thus 
implementing $U_{cond}$. 

Let every basic state be partitioned as:
\newline $e_i =|\mbox{place for code}\ [C], R_{\bar 1} ,R_{\bar 2} ,\ldots , 
R_{\bar l} 
\rangle$, where each register $R_{\bar i}$ in its turn is partitioned into a
place for argument, places for time instants and places for the corresponding 
frequencies. Here a complex index $\bar i$ contains one or two integers so 
that the length of $e_i$ is polynomial of $c$ and $n$ of at most second 
degree.

\section{Obtaining new algorithms from basic quantum tricks}

\subsection{GSA and amplitude amplification}

GSA proposed in (\cite{Gr}) is one of two 
basic 
quantum tricks. It is meant for quick getting of a quantum state $\bar a$
given the inversion along this state $I_{\bar a}$. An inversion along some 
state $\bar a$ is defined by 
$$
I_{\bar a} |\bar x\rangle =\left\{
\begin{array}{cc}
|\bar x\rangle ,\ &\mbox{if}\ x\bot a,\\
-|\bar a\rangle ,\ &\mbox{if}\ x=a.
\end{array}
\right.
$$
We also assume that $I_{\bar a}$ acts like identity if $\bar a$ does not
exist. A typical situation is when a state is unknown but the inversion along 
it can
be fulfilled easily. Say, let $\bar a$ be a solution of equation $f(x)=1$ with
a simply computable Boolean function $f$. Then the inversion $I_{\bar a}$ can 
be implemented by addition modulo 2 of $f(x)$ to an ancillary qubit
initialized by $\frac{|0\rangle -|1\rangle}{\sqrt{2}}$. This transformation
maps a state $|x, \frac{|0\rangle -|1\rangle}{\sqrt{2}} \rangle$ to the same 
state with the sign + or $-$ subject to the satisfaction of equality $f(x)=1$.
The transformation is unitary and can be easily fulfilled given a device
 fulfilling $f$. All sequential transformations in our formulas will be applied 
from right to left.

GSA is sequential applications of the transformation
$G=I_{\bar a}I_{\tilde 0}$ to a state $\tilde 0$ which is chosen 
randomly beforehand. If we apply this transformation $O(\sqrt{N})$ times where
$N$ is the
 dimension of main space then an observation of quantum part yields $\bar a$
 with visible probability whereas without quantum computer we would be
 compelled to spend
of order $N$ steps to find $\bar a$. 

A little difficulty here is that we don't know exactly a time instant $t$ when 
to stop
 iterations to make probability of error negligible that is needed when 
applying GSA as subroutine. Here the following simple trick helps.

Define a number $B=B(N)$ so that $1/B$ is an average value of $|\langle a\ |\ \tilde 0\rangle |$ for 
$\tilde 0$ uniformly distributed on a sphere of radius 1 in the space of inputs.
 A straightforward calculation shows that $B=O(\sqrt{N})$. Let $\GenArg_j$ be operators generating
arbitrary vectors $\bar a_j$ from the space of inputs belonging to independent
uniform distributions, $j\in\{ 1,2,\ldots ,k\}$, and let $\GenTimeArg_j$ be
operators generating time instants $t_j$ from independent uniform
 distributions on integers from the segment $[0,B]$. Arrange $k$ copies of two 
working
registers: for input and for storage of a time instant and fulfill the
corresponding operator $(I_{\tilde 0}I_{\bar a} )^{t_j}\GenArg_j \GenTime_j$
on each register. Now if $\bar a$ exists then the probability to obtain $\bar 
a$ 
observing any one register is at least 1/4 (it is shown in \cite{BBHT}) and 
the probability to obtain
any fixed other state will be negligible because our operators $\GenArg_j$
 generate independent uniformly distributed samples. If $\bar a$ does not 
exist
which means that $I_{\bar a}$ is identity then the probability to obtain any
fixed state will be negligible. Denote by $\bar a_j$ the contents of $j$-th 
register for argument in the resulting state. Consider the following 
criterion: if at least one fifth of $\bar a_j ,\ j=1,2,\ldots , k$ coincide
then we decide that $\bar a$ is this value, if not then $\bar a$ does not 
exist. 
Calculate
the error probability of this criterion. Let $K$ be a number of such $j$ 
that $\bar a_j =\bar a$. By the central limit theorem the probability that a 
fraction
 $\frac{(k/4)-K}{\sqrt{(k/4) \cdot (3/4)}}$ belongs to the segment $[\a_1 ,
\a_2 ]$
 is closed to 
$\frac{1}{\sqrt{2\pi}}\int\limits_{\a_1}^{\a_2}e^{-\frac{x^2}{2}} dx$.
Then the straightforward calculations give that the probability of that
$K\leq k/5$ will be of order $\int\limits_{\a_1}^{\infty}e^{-\frac{x^2}{2}}dx$ 
for $\a_1$ of order $\sqrt{k}$. Thus to make error probability of order
 $1/\sqrt{N}$ it would suffice to choose $k$ of order $n=\log N$. This 
method can be used not only for GSA but also for other algorithms. If
 a probability to obtain a right result for each of $k$ registers is some
 positive $p$ which does not depend on the dimensionality then to make this
probability $1/N_1$ it is sufficient to choose $k$ of order $\log N_1$. In
what 
follows we shall use this simple trick without special mentioning and will 
mark the simultaneous operations of the same kind on all working registers by
 $\bigotimes\limits_j $. We assume that all ensembles generated by the 
different $j$-th copies of operators are taken from the independent 
distributions.

We shall use the standard norm on operators in Hilbert space defined by 
$\| A\| =\sup\limits_{\| \bar x\| =1} \| A\bar x\|$. Given an operator $A$
denote by $A_\e$ such operator $A'$ that $\| A-A' \|\leq\e$. In what follows
 we shall use a simple trick described above making necessary copies of 
registers and so will 
raise an accuracy of our operators up to required level. When we must repeat 
an operator $T$ times the required accuracy of one application must be $1/T$ 
and it may be ensured by only linear price in memory as it was shown above.
This means that we shall always use $A_\e$ instead of $A$ without special 
mentioning where $\e =O(1/T)$ if an operator $A$ must be repeated $T$ times.

\subsection{Revealing of eigenvalues}

The second basic quantum trick is designed to reveal eigenvalues of a given 
unitary operator $U$. We shall define an operator 
revealing frequencies in accordance to
the work \cite{AL}.

Let $M=2^m ,L=2^p$. We are going to determine frequencies of unitary operators in
within $1/L$ where $L$ is a number of 
applications 
of
 $U$ required for the revelation of frequencies with this accuracy which means
that $1/M$ is an accuracy that is sufficient to tell apart eigenvalues of $U$. For the recognition of eigenvalues we put $p=m+4$ so that $L=16M$.

Denote by $(0.l)_p$ a number from 
$[0,1)$ of the form $l/L$. Let an operator $U$ have eigenvalues $e^{2\pi i \w_k}$ where frequencies $\w_0 , \w_1 , \ldots ,\w_{N'-1}$ are the different real numbers from $[0,1)$. Denote by $E_k$ the space spanned by all eigenvectors corresponding to $\w_k$. An arbitrary vector with the length 1 from $E_k$ will be denoted by $\Phi_k$. Thus every state $\xi$ has the
form
 $\xi =\sum\limits_{k=0}^{N'-1} x_k \Phi_k$. 

Let $\Omega =\{\tilde\w_{k,i} \}$ be some set of integers from $\{ 0,1, \ldots ,L-1\}$, $0\leq i\leq {\cal 
M}-1,\ 
0\leq k\leq N'-1$; $\ep ,\d >0$. Denote
$L^k_\ep (\Omega ) = \{ i:\  |(0.\tilde\w_{k,i} )_p - \w_k |\leq\ep$
or \newline
$|(0.\tilde\w_{k,i} )_p - \w_k -1|\leq\ep\}$. 

{\bf Definition 1}
{\it A transformation $W$ of the form 
$$
W:\ |\xi , 0^{m+4} \rangle \ar\sum\limits_{k=0}^{N'-1}
\sum\limits_{i=0}^{L-1}
\la_{i,k} |\Phi_k ,
\tilde\w_{k,i} \rangle
$$
is called a transformation of $W_{\d, \ep}$ type if 
for all $k$ and $\xi$ $\ \ \sum\limits_{i\in L_\ep^k (\Omega )}
|\la_{i,k}  
|^2 \geq |x_k |^2 (1-2\d )\ \ 
$.}

Thus, $\d$ is an error probability of getting right frequencies $\w_k$ by observation of the second register, and
$\ep$ is a precision of frequencies approximations.

{\bf Definition 2}
{\it A unitary operator $R$ is called revealing frequencies of $U$ if $R$
belongs to the type $W_{\frac{1}{K},\frac{K}{L}}$ for any $K\in\{ 
1,2,\ldots
 ,L\}$.\footnote{in what follows we shall use this notion only with $K=16$.}}

The key here is the quantum version of Fourier transform defined by 
$$
\QFT_L :\ \ |s\rangle \ar \frac{1}{\sqrt{L}}\sum\limits_{l=0}^{L-1} 
e^{\frac{-2\pi i s l}{L}}|l\rangle
$$
We need also the following generalization $U_{seq}$ of operator $U_{cond}$:
$$
\U^L_{seq} |x,a\rangle = | \U^a x,a\rangle .
$$
 This is the result of $a$ sequential applications of $U$ to the main
register. To implement this operator by means of $U_{cond}$ fulfill the
following cycle. For integer counter $j$ altering from 1 to the maximal 
value $L-1$ of $a$ apply $U$ iff $j\leq a$. Then one cycle consists of
 $U_{cond}$ with a properly prepared controller and the resulting operator 
will be $U^L_{seq}$. 

Define an operator revealing frequencies by
$$
\Rev =\QFT_L \ \U^L_{seq} \ \QFT_L ,
$$
where quantum Fourier transforms are applied to the second register 
\footnote{the first $\QFT$ can be replaced by Walsh-Hadamard transform as in 
\cite{AL} because on zero ancilla it is equivalent}. In the 
work \cite{Oz} it was proved that $\Rev$ is a transformation revealing
frequencies. Now we need more. For the redistribution
of amplitudes $x_k$ we shall need also a transformation $\Rest$ cleaning the
 second register. An ideal candidate for this role would be $\Rev^{-1}$ but a
 problem is that it requires an application of $U^{-1}$ that is
 physically unrealizable given only device fulfilling $U$ excluding evident 
cases where say $U=U^{-1}$. We can use this easiest definition of $\Rest$ only
in case when we are given a circuit implementing $U$ (say, gate array) because 
then $U^{-1}$ is accessible for us as well as $U$. But if $C$ is given only as 
a black box then the restoring operator should be defined separately.

We shall find an operator restoring ancilla in the form 
$$
\Rest =\Rev D
$$
where $D$ is some operator of turning. 

Let we are given some integers $\tilde \w^L_k$ of the form $\frac{q}{L}$, $q$ - integer, 
$\tilde\w^L_k \approx\w_k$. Then we could define an operator of turning $D$ by $D|\Phi_k ,
l\rangle =e^{-2\pi i(L-1)\d_{k,l}}|\Phi_k ,l\rangle$ where $\d_{k,l}
=\tilde\w_k^L -(0.l)_m$. 
It was proved in \cite{Oz} that
$\| (\Rest\Rev |\chi ,\bar 0\rangle -|\chi ,\bar 0\rangle \|<7M/L$ which 
means that so defined restoring operator really restores zeroes in the second
 register after action of $\Rev$ provided $L$ is large enough. To create these good approximations we apply a bit more general construction. Put 
$$
D=\Enh\ \tilde D\ \Enh
$$
 where an operator $\Enh$ calculates an integer 
function 
$h(l)$ giving a good 
approximation $(0.h(l))_p$ of frequencies in within $1/L$ given their rough
approximation $(0.l)_m$ in within $1/M$ and places them into ancilla, 
$\tilde D$ turns 
each eigenvector on appropriate angle: \newline$\tilde D|\Phi_k \rangle =e^{-2\pi
i(M-1)((0.h(l))_p -(0.l)_p )}|\Phi_k$ and the last application of $\Enh$ 
cleans
ancilla. An operator $\Enh$ is accessible given good approximations of
 eigenvalues. Thus our operator $\Rest $ restores zeroes in ancilla in within 
$1/L$. 

We can reach an accuracy $1/L$ of all operators of type $\Rest$ that will be
 less than $1/t$ where $t$ is the number of all steps in computation and this 
accuracy can be guaranteed with $\log L=p$ registers. Emphasize that this 
difficulty with eigenvalue precision arises only when $U^{-1}$ is 
inaccessible - in the problem of recognition of electronic circuits in the 
section 3.4 where we must choose $L\gg M$. 

Operators $\Rev$ and $\Rest$ can be built in the form of quantum gate array 
using universal quantum Klini function $\App$ where a code $[C]$ of circuit 
generating $U$ is a part of input. We shall write the operator $U$ corresponding 
to these two operators as its upper index. 

\section{Problems of recognition}

\subsection{Getting eigenvectors and recognition of eigenvalues}

Our assumption about sparse spectrum now is stated as $L=16M=const$. In view of that $\Rev$ reveals frequencies it belongs to the type $W_{\frac{1}{16} ,\frac{1}{M}}$. By the definition of $W_{\d ,\ep}$ it means that $\Rev$ gives a state $\sum\limits_{k=0}^{N'-1}\sum\limits_{i=0}^{M-1} 
\la_{i,k} |\Phi_k ,
\tilde\w_{k,i} \rangle$ where the seven eighth of probability is concentrated  
on such pairs $i,k$ that $(0.\w_{i,k})_m$ is closed to $\w_k$. This
means that we can get eigenvalues with high probability observing the second
register whereas the first register will contain the corresponding
eigenvector. This way of getting eigenvectors was proposed in the papers
\cite{AL,TM}. The first disadvantage of it is irreversibility. Observing a 
state we lose full information about
it and cannot use this state again that is very important for building of 
nontrivial quantum algorithms. The second disadvantage is that this way gives 
a random eigenvector when it is typically required to obtain the eigenvector 
corresponding to a given frequency. 

Let we are given a good approximation $\tilde\w^L$ of some frequency $\w$ 
written as a string of
 $p$ its sequential binary figures and let ${\cal E}_{\w} =\{ 
\Phi^{\w}_1 , \ldots \Phi^{\w}_l \}$ be a basis of the subspace $E_{\w}$ 
of eigenvectors corresponding to all frequencies $\w' \approx\w$. We shall build
 an operator $\State_{\w}$ which 
concentrates the bulk of amplitude on some superposition of the corresponding
eigenvectors: $\sum\limits_{j=1}^l \la_j
\Phi^{\w}_j \in E_{\w}$. For this aim we are going to apply GSA. Let
$| \bar a\rangle$ be some randomly chosen vector from the main space:
$|\bar a\rangle =\sum\limits_{j=1}^l \mu_j \Phi^{\w}_j +\sum\limits_s \nu_s
\Phi_s$ where all eigenvectors from the second sum correspond to frequencies 
$\w'\not\approx\w$. Here our target state will be the following vector 
$E_{\w}(\bar a)=\sum\limits_{j=1}^l \la_j \Phi^{\w}_j$ where 
$\la_j=\frac{\mu_j}{\sqrt{\sum\limits_{j=1}^l |\mu_j |^2}}$, that is a vector 
of the length 1 directed along the projection of $\bar a$ to subspace $E_{\w}$. 

Let $A$ be some set of vectors. We denote by $I_A$ an operator changing the sign of all vectors from
 $A$ and remaining unchanged all vectors orthogonal to $A$. 
We need to obtain the operator $I_{E_{\w}}$ constrained to the two 
dimensional subspace $S(\bar a,\w )$ spanned
by vectors $|\bar a\rangle$ and $E_{\w} (\bar a)$. 

Let $\Rev_j ,\ \Rest_j$ be $j$-th copies of the operators $\Rev ,\ \Rest$
 acting on the corresponding places of $j$-th register. Denote by $l_j$ a
string contained in the place for frequency of $j$-th register. Put
$$
\tilde I_{E_\w} =\bigotimes\limits_j ^v \Rest_j \ \Sign_{\w} \ 
\bigotimes\limits_j^v \Rev_j
$$
then $\Sign_{\w}$ changes a sign if and only if for at least $1/2$ of all 
$j$ $|(0.l_j )_p -(0.\tilde\w^L )_p|\leq 1/L$.\ \footnote{We could choose any fixed $\rho :\ \frac{1}{8}<\rho <\frac{7}{8}$ instead
of $1/2$. Really, so defined $\tilde I_{E_{\w}}$ will change the sign of all
 $\bar a\in E_{\w}$. If $\bar a\bot E_{\w}$ then the probability to obtain
 $\w$ observing the frequency from $\Rev$ is less than $\frac{1}{8}$.}
Applying reasoning from the end of section 2.1 we conclude that 
the actions of $I_{E_\w}$ and $\tilde I_{E_\w}$ restricted on $S(a,\w )$ 
will differ on less than $\frac{1}{2^{O(v)} }$ and thus this different can be 
done very small by only linear growth of memory. We thus will omit $\tilde{}$ 
in our notations. 

We define
$$
\St =\GenArg^{-1} \GenTimeArg^{-1} (I_{\bar a}I_{E_{\w}})^t \GenTimeArg\ \GenArg
$$
where $\GenArg$ and $\GenTimeArg$ generate a pair $\bar a,\ [C]$ and a time 
instant $t$ correspondingly where 
$C$ is a gate array implementing $I_{\bar a}$. Here the actions of $I_{\bar a}$ 
are implemented by the
 universal function of application $\App$. Then the result $\xi=\St
|\bar 0\rangle$ of its action
on $\bar 0$ will be closed to $E_{\w}(\bar a)$. Really, $|\langle
E_{\w}(\bar a) |\xi\rangle |=|\sin (2t \arcsin\langle\bar a|E_{\w}(\bar a) 
\rangle )|$ (look at \cite{BBHT}). 
An average value of $|\langle\bar a|E_{\w}(\bar a)\rangle |$ with the uniformly 
distributed probability of choice $\bar a$ and $t$ over all space and time frame 
$[0,B]$ correspondingly will be of order $1/\sqrt{N}$. 
Thus if $t$ is chosen randomly from the uniform distribution over
$1,2,\ldots ,B$ then $|\langle E_{\w}(\bar a) |\xi\rangle |^2$ will have
average value not less than $1/4$. Of course it would be much more convenient
to obtain $E_{\w}(\bar a)$ with the error probability converging to zero that 
is possible by the method described in section 2.1. Namely, arrange $h$ equal 
registers for the states
$\chi_k , \ k=1,2,\ldots , h$ in the main space, the corresponding $h$
registers for frequencies and associate the variable $t_k$ with each $k$-th
register. 
Let $\St_k$ be a pattern of $\St$ operator
acting on $k$-th register. Remind that operators $\GenArg_k$ and
 $\GenTimeArg_k$ generate independent distributions for different
 $k=1,2,\ldots ,h$. Now we define
\begin{equation}
\State^{\w} =\St_1\bigotimes\St_2
\bigotimes\ldots\bigotimes\St_h .
\label{State}
\end{equation}
So defined operator being applied to zero initial state gives a state
 $\chi_1 \bigotimes\chi_2 \bigotimes\ldots\bigotimes\chi_h$ where an average value of $\ |\langle E_{\w}(\chi_k ) |\chi_k \rangle |^2$ will be closed to some number not less than $1/4$ with
the vanishing probability of error. By the way it means that if we then
apply to this state the corresponding operators revealing frequencies:
$\Rev_1 \bigotimes\Rev_2 \bigotimes\ldots\bigotimes\Rev_h$ then in the
resulting state $\chi$ the majority of amplitude will be concentrated on
such basic states for which at least $\frac{5}{32}$ of all registers for
frequencies contain numbers $l$ for which $|(0.l)_m -(0.\tilde\w^L )_p |<1/L$.\ \footnote{Note that in this criterion $\frac{5}{32}$ could be replaced by any 
$\rho:$
 $0<\rho <\frac{1}{4}\cdot\frac{7}{8}=\frac{7}{32}$.}. 
On the other hand, if $\w$ is not a frequency at all then the probability
to obtain such basic state will be vanishing in view of independence of 
distributions generated by $\GenTimeArg_k$ and $\GenArg_k$ for the different 
$k$. 

The time complexity of this algorithm is of order $M\sqrt{N}\cdot n^2$. The 
last multiplier arises due to the copying of registers.
Thus we have a solution of the first problem of recognition of eigenvalues.

\subsection{Finding thermodynamic functions}

Let we are given a structure of molecule of a gas. The problem is to find its 
thermodynamic function (\ref{Therm}). In view of that a summand in this sum 
quickly converges to zero it is sufficient to find few first summands. Thus 
it is sufficient to be able to find degree of degeneracy of the subspace 
corresponding to frequencies $\w'\approx\w$ for any $\w =l/M$. Let $E_0 
<E_1 <\ldots <E_s$ be energy levels of a molecule or eigenvalues of its 
Hamiltonian $H$. Then the operator of evolution in time frame $t$ will 
be $U=e^{-\frac{iH}{h} t}$. An addition of a diagonal matrix $r\cdot I$
 with constant $r$ to a Hamiltonian does not change the physical picture. 
Then choosing $r=-E_s ,\ t=\frac{h}{2\pi E_s}$ we obtain a unitary operator 
$U$ which frequencies belong to the segment $[0,1)$ and thus we reduce the 
problem to the case we have already dealt with. 

Assume that $M$ is fixed and we must examine only few frequencies closed 
to $0$. At first we can recognize all numbers of the form $l/M$ that are 
frequencies in within $1/L$. Let $\w$ be such number. Show how we can find
 the degree of degeneracy $d$ of the corresponding subspace. This is the 
dimension of subspace $E_{\w}$ spanned by eigenvectors corresponding to
frequencies $\w'\approx\w$. Our plan is the following. Build an operator
 $I_{E_{\w}}$ of reflection along this subspace. Then using a counting 
procedure built in the paper \cite{BBHT} we evaluate the time required 
for turning of arbitrary initial vector till this subspace. This time 
will be about $\sqrt{\frac{N}{d}}$ and thus we find $d$. Fix some $\e >0$ 
and show how to obtain the value of $d$ in within $\e d$. 

Let operators $\GenTimeArg^a_j$ generate time instants $t_j$ from independent
 uniform distributions on the segment $[0,[a]]$, where $a$ is nonnegative 
number. For $a$ from 1 to $\sqrt{N}$ fulfil the following loop of three steps:

\begin{enumerate}
\item Apply an operator 
$$
\bigotimes\limits_j \left[\bigotimes\limits_k \Rev_{j,k}\right] (I_{\bar a}
I_{E_{\w}})^{t_j}\GenTimeArg^a_j \GenArg_j
$$

\item Find the fidelity of result that is the number of all $j$ for 
which at least $\frac{7}{8}-\e$ of all $k$ are such that $\w_{j,k}
\approx\w$. If the fidelity of this step is larger than on the previous, 
then proceed the loop, if not then stop.

\item Replace $a$ by $4a/3$.
\end{enumerate}
If by the point 2 we finish computation then the current value $a$ is taken as 
the rough approximation of $d$ from above. We have $3a/4\leq d\leq a$. To 
find $d$ more exactly divide the segment $[3a/4,a]$ to $[1/\e ]$ equal parts
 by points $a_0 <a_1 <\ldots <a_l$ and repeat the procedure from above 
sequentially for all $a_i$. Thus we shall determine $d$ in within $g(\e )d$ 
where function $g$ quickly converges to zero with $\e$. Thus our algorithm 
finds $d$ and thermodynamic functions with arbitrary relative error in the 
time $O(\sqrt{N})M$ where the constant depends on admissible error. The more 
refined algorithm can be obtained if we apply the method of counting from
 the work \cite{BHT}. In that work quantum Fourier transform is used analogously
 to Abrams and Lloyd operator $\Rev$ only in order to find a time period 
of function $G|\xi ,t\rangle =|G^t \xi ,t\rangle$ that is about $\sqrt{N/d}$.
 Their method gives the accuracy of order $\sqrt{d}$ which means that the 
relative error converges to zero if $d\ar\infty$. 

\subsection{Recognition of molecular structures}

Now take up a problem of recognition of molecular structures. Here we are given
a spectrum of molecule and a problem is to recognize its construction. Note
that now we have not access to a device but it is sufficient to find an 
arbitrary device generating this spectrum. Clarify the formulation assuming the following form of determining spectrum. Let we are given a set 
$\bar w=\{ w_1 ,\ldots ,w_Q \}$ of numbers from $[0,1)$ of the form $w_i =
\frac{l_i}{M}$ each where $l_i \in\{ 0,1,\ldots ,M-1\}$. Denote by $F$ a 
subspace spanned by vectors of the form $|l_i \rangle ,\ i=1,\ldots Q$. A 
spectrum $S$ is determined by this set $\bar w$ if 
\begin{itemize}
\item[a)] for each $\w\in S$ there exists its good approximation $w_i \in\bar w$: 
$|w_i -\w |\leq\frac{1}{L}$ and
\item[b)] each $w_i \in\bar w$ is a good approximation of some $\w\in S$.
\end{itemize}
We would obtain a little different formulations of the problem if we
want to find a circuit whose spectrum only contains one given set of 
frequencies and/or does not contain some other set, or permit some more general form of sparse set for $\bar w$ 
instead of $\frac{l_i}{M}$. These versions of the problem at hand have 
similar solutions. 

As above we shall find a recognizing algorithm in GSA form
\begin{equation}
(I_{\tilde 0}I_{cir,\ \bar w})^t
\label{GSA}
\end{equation}
where $\tilde 0$ is arbitrarily chosen vector from the space spanned by codes
 of circuits, $t=O(\sqrt {T})$ where $T$ is a number of all possible circuits, 
and $I_{cir, \bar w}$ is reflection along all
 such codes $[C]$ that $\Spectr (U_C )$ is determined by $\bar w$. Now it is 
sufficient to build $I_{cir,\ \bar w}$. 

Choose $B_f =O(\sqrt{Q})$ so that a 
randomly chosen vector $w\in F$ satisfies $|\langle w|w_1 \rangle |>1/B_f 
$ with probability $0.99$. Let $\GenFreq_j$,
$\GenTimeFreq_j$ be operators generating correspondingly: 
a linear combination of frequencies $\tilde\w_j \in F$ and time instant $t_{freq,\ j} \leq B_f$ - all
 these objects from the corresponding uniform distributions over all possible
values, and a code of gate array generating inversion along the corresponding state $\tilde \w_j$. These operators will generate the objects in the corresponding ancillary registers. Denote by $\w_j$ a frequency contained in the $j$-th register (that is initially $\tilde\w_j$).

Assume that a code of circuit generating $U$ is fixed. Define an operator 
$I_{cir,\ \bar w}$ by 
$$
\begin{array}{l}
I_{cir,\ \bar w}=\bigotimes\limits_j \left[\GenFreq_j^{-1}\GenTimeFreq_j^{-1}
(I_{BadFreq,\ \bar w ,j}I_{\tilde\w_j})^{t_{freq,\ j}}\right]\SignGoodFreq\\
\qquad {}\bigotimes\limits_j 
\left[ (I_{\tilde\w_j}I_{BadFreq,\ \bar w ,j})^{t_{freq,\ j}}
 \GenFreq_j 
\GenTimeFreq_j \right] .
\end{array}
$$
where $I_{BadFreq,\ \bar\w ,j}$ will invert a sign of states with
 "bad frequencies" in $j$-th register that are such values of $\w_j$ of the 
form $\frac{l}{M}$,
$l\in\{ 0,1,\ldots ,M-1\}$ which either belong to $\bar w$ and are not a good 
approximation of frequencies $\w\in \Spectr (V)$ or do not belong to $\bar w$ 
but have a
 closed $\w\in \Spectr (V)$: $|\w_j -\w |\leq\frac{1}{L}$; for all other 
frequencies this operator acts like identity. Application of the sequence 
preceding $\SignGoodFreq$ concentrates amplitude on "bad frequencies". 
Note that $I_{\tilde \w_j}$ can be implemented by a given code by means 
of quantum Klini operator $\App$. The 
following application of $\SignGoodFreq$ inverts a sign of state subject to 
are there bad frequencies or not. Namely, for codes $[C]$ without bad 
frequencies $\SignGoodFreq$ changes the sign, for codes $[C]$ with bad
 frequencies it makes nothing. The
 following operators clean all ancilla. Thus so defined $I_{cir,\ \bar w}$
 will invert a sign of exactly those codes $C$ for which $\Spectr(U_C )$ 
is determined by $\bar w$. We need to define two types of operators: 
$\SignGoodFreq$ and $I_{BadFreq,\ \bar w ,j}$.

Associate with each $\w_j$ contained in $j$-th register the family of 
registers enumerated by two indices $j,k$ and containing frequencies 
$\w_{j,k}$. 

{\bf Definition 3}. {\it Call a family of all $\w_{j,k}$ good if for at 
least $1/5$ from all $j$ the following property takes place: for at least 
$1/10$ of all $k$ $\ \w_{j,k}\approx \w_j \in \bar w$.}

 Registers enumerated by different $k$ for a fixed $j$ are designed for the 
application of $j$-th copy of operator 
$\State^{\w}$ defined in the previous section. Here it has the form 
$\State^{\w_j}$. Each $k$ corresponds to the operator $\St_k$ from the
 definition (\ref{State}) so that each $\w_{j,k}$ will be a frequency 
obtained from the result of $\St_k$ action. 

At first build an operator $I_{BadFreq, \bar w ,j}$. Put
$$
I_{BadFreq, \bar w ,j}=\bigotimes\limits_{j,k} 
\left[ {(\State^{\w_j})}^{-1} \Rest_{j,k}\right]\Sign' 
\bigotimes\limits_{j,k}\left[\Rev_{j,k}\State^{\w_j}\right]
$$
where an operator $\Sign'$ will change a sign of only states with bad 
families of frequencies. 

If a frequency $\w_j$ is bad then in the previous section it was shown that 
only for vanishing part of all $k$ we can have $\w_{j,k}\approx\w_j \in\bar w$ 
and before $\Sign'$ almost all probability will be concentrated on bad 
families $\w_{j,k}$, hence $I_{BadFreq, \bar w, j}$ changes the sign.

If $\w_j$ is good then it belongs to $\bar w$ and has a closed $\w' \in S$.
 By the previous section for about $\frac{7}{8}\cdot\frac{1}{4} =
\frac{7}{32} >\frac{1}{5}$ of all $k$ will be $\w_{j,k}\approx\w\in\bar w$
 and before $\Sign'$ almost all probability will be concentrated on good 
families, hence the sign will not be changed. 

Thus $I_{BadFreq ,\bar w ,j}$ is defined correctly.

Put 
$$
\SignGoodFreq =\bigotimes\limits_{j,k} \left[
{(\State^{\w_j})}^{-1}\Rest_{j,k}\right]\Sign 
\bigotimes\limits_{j,k} \left[\Rev_{j,k}
\State^{\w_j} \right]
$$
where an operator $\Sign$ changes a sign only for states with good families of 
frequencies. If a frequency $\w_j$ is not bad then for about 
$\frac{1}{4}\cdot\frac{7}{8} =\frac{7}{32}$ of all $k$ 
will be $\w_{j,k}\approx\w_j \in\bar w$. If a frequency $\w_j$ is bad then we 
can obtain $\w_{j,k}\approx\w_j \in\bar w$ only for the vanishing part of
 $k$ as it was shown in the previous section.
Thus $\SignGoodFreq$ acts how it is needed\footnote{Again we could 
take arbitrary $\rho_1 :\ 0<\rho_1 <1$ instead of $\frac{1}{10}$ and $\rho_2
:\ 0<\rho_2 <\frac{7}{32}$ instead of $1/5$ in the definition of a good 
family.}.

Now calculate the complexity of our algorithm recognizing molecular circuit. 
The first multiplier $\sqrt{T}$ issues immediately from (\ref{GSA}). The 
following multiplier as $\sqrt{Q}$ issues 
from the immediate definition of $I_{cir,\ \bar w}$. At last the definition of 
$I_{BadFreq,\ \bar w}$ brings the multiplier $M\sqrt{N}$. The
 resulting complexity will be of order $M\sqrt{TNQ}n^2$. 

\subsection{Distinguishing of eigenvectors of two operators with the same 
eigenvalue}

Now we are going to take up the most difficult of our problems - a problem of 
recognition of electronic devices. The difficulty is that here we need not 
to find a circuit with given spectrum but to simulate an action of given 
circuit. Remind that now we assume that 
frequencies can be determined in within $1/L$ given their approximation in
within $1/M$ where $L\gg M$. 

As a first step we consider the following question. Given two operators $U$ 
and $V$ having the
same eigenvalue $\w$ how to find a difference between the corresponding
eigenvectors? Let $L_{\w}^U ,\ L_{\w}^V$ be the subspaces spanned by 
eigenvectors of $U$ and $V$ corresponding to all frequencies $\w'\approx\w$.
 (A particular case is when $\w$ is a frequency of $U$ but not of $V$. Here 
$L_{\w}^V =\emptyset$ and our algorithm will work in this situation.) We 
shall omit the index $\w$ in the notations. For $u\in L^U ,\ \| u\| =1$ put
$$
\mu_u =\min\{\sqrt{1-|\langle u|v\rangle |^2}\ |\ v\in L^V ,\ \| v\| =1\} .
$$
that is the sine of angle between a vector $u$ and the subspace $L^V$ or a 
distance between $u$ and this subspace, and analogously define $\mu_v$ for
$v\in L^V ,\ \| v\| =1$.

Put 
$$
\mu_U =\max\limits_{u\in U} \mu_u ,\ \mu_V =\max\limits_{v\in V} \mu_v .
$$
Then say $\mu_U =0$ means that $U\subseteq V$. If the dimension of spaces
 $L^U ,L^V$ are equal then $\mu_U =\mu_V$, if they are not equal, say $\dim 
L^U >\dim L^V$ then $\mu_U =1$. 

Let $d=d(N)$ be some function taking values from $(0,1]$. 

Call these subspaces $d$- distinguishable if some of $\mu_U ,\mu_V$ is not less 
than $d$, or some of the subspaces is empty and another is nonempty. 

We shall build a procedure that determines are these subspaces the same or not 
provided they can be either $d$- distinguishable or 
coincident. The less the function $d(N)$ is the more accurate our recognition 
will be. Let $L^U \cap L^V =L_0$. Then $L^U =L_0 \bigoplus L'_U$ and $L^V 
=L_0 \bigoplus L'_V$. Note that if $L'_U \neq\emptyset$ then for all vectors 
from $L'_U$ of the length 1 their distances from $L^V$ are exactly $\mu_U$ and 
the same thing takes place with $L^V$ if $L'_V$ is not empty. Let $L'$ be linear subspace spanned by vectors from $L'_V \cup L'_U$. Denote by $\Proj_A B$ a projection of
subspace $B$ to subspace $A$. If $\dim L^U >\dim L^V$ then we have the 
following expansion to the sum of orthogonal subspaces: $L^U =L''_U 
\bigoplus\Proj_{L^U} L^V$, where $L''_U$ is a subspace in $L^U$ 
consisting of vectors orthogonal to $L^V$. Let $L''_V$ be defined
symmetrically. 

Then either 
\begin{itemize}
\item $L^U =L^V$ or 
\item $\dim\ L^U =\dim\ L^V$ and $L' \neq\emptyset$, or
\item $\dim\ L^U >\dim\ L^V$ and $L''_U \neq\emptyset$, or
\item $\dim\ L^U <\dim\ L^V$ and $L''_V \neq\emptyset$.
\end{itemize}

We define the main operator determining the equality of $L^U$ and $L^V$ by 
\begin{equation}
\begin{array}{l}
\Difference =\Differ ^{-1}\ \SignDif\ \Differ ,\\
\qquad {} \Differ =\Dif_{same\ dim} \Dif_{L^U >L^V} \Dif_{L^U <L^V} 
\Dif_{L^U >L^V}^{ort} \Dif_{L^U <L^V}^{ort}
\end{array}
\label{Difference}
\end{equation}
where $\SignDif$ changes a sign of main ancilla $\a_{dif}$ iff at 
least one ancilla from the list $\bar\a =\{ \a_{same\ dim} ,\a_{L^U >L^V} ,
\a_{L^U <L^V} ,\a_{L^U >L^V}^{ort} ,\a_{L^U <L^V}^{ort}\}$ contains 1, and 
each operator of the sort $\Dif$ changes the corresponding ancilla from $\bar 
\a$ in cases
\begin{itemize}
\item $\dim\ L^U =\dim L^V$ and $L^U \neq L^V$,
\item $\dim L^U >\dim L^V$ and $\mu_V <\sqrt{2/3}$,
\item $\dim L^U <\dim L^V$ and $\mu_U <\sqrt{2/3}$,
\item $\dim L^U >\dim L^V$ and $\mu_V >\sqrt{1/3}$, or $L^V =\emptyset$,
\item$\dim L^U <\dim L^V$ and $\mu_U >\sqrt{1/3}$, or $L^U =\emptyset$
\end{itemize}
 correspondingly and all of them do
 nothing if $L^U =L^V$. In view of symmetry it is sufficient to define $\Dif$ 
operators in the first, second and fourth cases. Note that the first case
 $\dim L^U =\dim L^V$ is the only non-degenerate case and the definition of 
$\Dif$ here will be the most difficult. 

\nn
{\bf Definition of } $\Dif_{same\ dim}$.
\n

Suppose that $\dim L^U =\dim L^V$. 
Our first aim now is to build an operator $\Inv$ which acts as identity if
$L^U$ and $L^V$ are coincident and acts like $I_{L'}$ if they are 
$d$- distinguishable. Arrange the first two ancillary qubits $\a_U , \a_V$
which will signal that a state at hand has a projection at least of the length 
$1/3$ to $L^U$ or, correspondingly to $L^V$. Consider the following operator 
$$
\Check = \bigotimes_s \Rest^V_s \Anc_V \bigotimes_s \Rev^V_s \bigotimes_s
 \Rest^U_s \Anc_U \bigotimes_s \Rev^U_s
$$ where $\Anc$
inverts the corresponding ancilla if and only if at least nine tenth of copies
 for the respective frequencies are equal to $\w$ in within $1/M$. It coincides 
with the inverse operator $\Check^{-1}$.

Let $t$ be some random integer from the segment $[0,\left[\frac{2}{d}\right] ]$. 
We define the following operator of Grover's type:

\begin{equation}
\Turn_t =(I_{L^U} I_{L^V})^t
\label{Turn}
\end{equation}
Call two subspaces $L^U$ and $L^V$ almost orthogonal iff for some 
$\mu\in\{\mu_U , \mu_V \}$ $\sqrt{1-\mu^2}\leq 1/30$. If $L^U$ and $L^V$ are not almost orthogonal then given
some $a\in L'_U$ ($a\in L'_V$) an average distance between $\Turn_t |a\rangle$ 
and $L_U$ ($L_V$) will be at 
least 1/2 if $L^U$ and $L^V$ are $d$- distinguishable and zero
 if these subspaces are coincident. To tell apart close location and almost orthogonality build two operators: $\Dist_{ort}$ and $\Dist_{closed}$. 

At first suppose that $L^U$ and $L^V$ are almost orthogonal. Then $\a_U =1$ 
means that $\a_V =0$. Introduce a notation
$$
L(\a_U ,\a_V )=\left\{
\begin{array}{cc}
L^V ,&\ \mbox{if}\ \a_U =1 ,\\
L^U ,&\ \mbox{if}\ \a_V =1.
\end{array}
\right.
$$
Let $\bar a$ be a vector at hand from the space of inputs. Note that if $L^U 
\neq L^V$ then for each $\bar a\bot L'$ we shall have $\a_U =\a_V$ because
 such $\bar a$ belongs to the subspace spanned by $L_0$ and orthogonal 
subspace to $L^U \cup L^V$. 
The first operator $\Dist_{ort}$ will do nothing if $\a_U =\a_V$ and will
 change a sign and a special ancilla $\a_{ort}$ if a projection of $\bar a$
 to $L(\a_U ,\a_V )$ is less than $1/30$. 

The second operator $\Dist_{closed}$ will act like identity if $\a_U =\a_V$ and 
will change
 a sign in case when the following conditions are satisfied simultaneously: 
$\bar a\in L'$, $L^U$ and $L^V$ are distinguishable, $\a_{ort} =0$. 

Put 
$$
\Dist_{ort} =\bigotimes_j \Res_j \Si_{\neq\w}\bigotimes_j \Re_j
$$
where $\Re$ ($\Res$) denotes $\Rev^V$ ($\Rest^V$) if $\a_U =1,\a_V =0$, 
$\Rev^U$ ($\Rest^U$) if $\a_V =1,\a_U =0$, and identity if $\a_U =\a_V$;
$\Si_{\neq\w}$ changes a sign inverting simultaneously
$\a_{ort}$ iff at least a half of frequencies $\w_j$ are such that $|\w_j -\w 
|>1/M$ and $\a_U \neq\a_V$. If we want to clean the second ancilla after the
action of $\Dist_{ort}$ and remain change in sign then we can use an operator 
$\Dist^{-}_{ort} =\bigotimes_j \Res_j S_{\neq\w}\bigotimes_j 
\Re_j$ where $S$ acts like $\Si$ only without changing a sign. 

The second operator will be defined by the following equations
$$
\begin{array}{l}
\Dist_{closed}=D_1^{-1} \ldots D_n^{-1}S'D_n D_{n-1}\ldots D_1 ,\\
\qquad{} D_j =(\GenTimeArg_j )^{-1}(\Turn_{t_j}^j )^{-1}\left[\bigotimes\limits_k 
\Rest^U_{j,k}\right] \Sig^j_{\neq\w}\left[\bigotimes\limits_k \Rev^U_{j,k}\right]\Turn_{t_j}^j
 \GenTimeArg_j\\
\qquad{} j=1,2,\ldots ,n,
\end{array}
$$
where operator $\Sig^j_{\neq\w}$ changes the corresponding ancilla $\beta_j$ 
only in one of the two cases:
\begin{itemize}
\item[1)] $\a_U =1$ and at least a half of $\w_{j,k}$ are such that $|\w_{j,k} -\w 
|\geq 1/M$, or
\item[2)] $\a_U =0,\ \a_V =1$ and at least a half of $\w_{j,k}$ are such that 
$|\w_{j,k} -\w |<1/M$.
\end{itemize}
 An operator $S'$ changes a sign iff some of $\a_U ,\a_V$ is nonzero and at 
least $1/20$ of all $\beta_j$ contain 1. 

Consider the action of $\Dist_{closed}$ following to $\Check$ on an input
 vector $\bar a$. Let at first $L^U \neq L^V$ which means that they are 
distinguishable. 

If $\bar a\bot\ L^U,L^V$ then $\a_U =\a_V =0$ and $\Dist_{closed}$ makes 
nothing.

If $\bar a\in L_0$ then $\a_U =\a_V =1$ and all $\Sig^j_{\neq\w}$ makes 
nothing 
because for almost all $j$ about $3/4$ of $\w_{j,k}$ are closed to $\w$:
$|\w_{j,k} -\w |\leq 1/M$, hence $S'$ and $\Dist_{closed}$ do nothing. 

Let $\bar a\in L'$. Prove that $\Dist_{closed}$ changes a sign. Expand $L'$ 
to the sum of orthogonal subspaces: $L'=L'_U \bigoplus {L'_U}^{ort}$. Denote the
result of action of $\Turn^j_{t_j}$ on $\bar a$ by $\bar a_j$. 

If $\a\in L'_U$ then $\a_U =1$, and for more than $1/10$ of all $\bar a_j$ 
revealed frequencies are not closed to $\w$ with probability about 
$\frac{3}{4}\cdot\frac{9}{10}$, hence a sign will be changed by the point 1).

If $\bar a\in {L'_U}^{ort}$ then by the same reason we obtain the change of 
sign by the point 2).Hence $\Dist_{closed}$ changes a sign for all $\bar a\in 
L'$. 

Now we can define $\Inv$:
$$
\Inv =\Check\ \Dist^{-}_{ort} \Dist_{closed}\Dist_{ort}\ \Check .
$$

For $a\bot\ L^U,L^V$ we have $\Inv |a\rangle =|a\rangle$ because $\Check$ gives 
zero in
 ancilla $\a_U ,\a_V$ thus depriving the following operators ability to change 
somehow a state vector. If $a\in L_0$ then $\Inv |a\rangle =|a\rangle$ because 
$\Dist_{ort}$ makes nothing and $\Dist_{closed}$ makes nothing as well. Thus 
$\Inv |a\rangle =|a\rangle$ for $\bar a\bot L'$, and 
$\Inv |a\rangle =-|a\rangle$ for $a\in L'$. 

Now we are ready to build an operator $\Dif_{same\ dim}$ inverting the
ancilla $\a_{same\ dim}$ if and only if $L^U$ and $L^V$ are distinguishable. 
Let $\Gen$ 
generate a list $y, [I_y ] , [C_Z]$ where $[C_Z]$ is a code of circuit 
generating some unitary 
operator $Z=Z^{-1}$ having only eigenvalues $1$ and $-1$ that is its 
frequencies are $0$ and $1/2$ and the space corresponding to frequency $0$
 is one dimensional where $y$ is its basic vector. As usually index $j$ means 
that the corresponding vectors $y_j$ are taken from the uniform distribution 
on all possible vectors. Assume that operators of the form $\Gen^{-1}$ are 
accessible for us as well. Put
\begin{equation}
\begin{array}{l}
\Dif_{same\ dim} =\bigotimes\limits_j \left[\GenTimeArg_j^{-1}\Gen_j^{-1}  
(\Inv_j \ I_{y_j} )^{t_j}
\Rest^{Z_j}_j
\right]\Change\\
\qquad {}\bigotimes\limits_j \left[\Rev^{Z_j}_j
(I_{y_j} \ \Inv_j )^{t_j}\Gen_j \GenTimeArg_j \right]
\end{array}
\label{Dif}
\end{equation}
where each copy of $\Inv$ acts on the register where initially is placed 
$y_j$ , $\Change$ makes a desired change in a resulting qubit $\a_{same\ dim}$
 provided at least $5/32$ of all frequencies differ from $0$ in more than 
$1/M$. 

The group $(I_{y_j} \ \Inv_j )^{t_j}$ of GSA type turns essentially a vector 
$y_j$ generated by $\Gen_j$ if and only if $L^U$ and $L^V$ are $d$- 
distinguishable. 

If $L^U =L^V$ then $y_j$ remains unchanged and at least $7/8$ of all frequencies 
will be closed to 0.

If $L^U \neq L^V$ then for the result of the turn of $y_j$ at least 
$\frac{7}{8}\cdot\frac{1}{4}=\frac{7}{32}$ of frequencies will be far from 0 
because 
they must be closed to $1/2$.\ \footnote{Thus we could take any number $\rho :\ 
\frac{1}{8}<\rho <\frac{7}{32}$ instead of $\frac{5}{32}$ in the definition of 
$\Change$.}

\nn
{\bf Definition of} $\Dif_{L^U >L^V}$.
\n

Suppose that $\dim L^U >\dim L^V$ and $\mu_V <\sqrt{2/3}$. Remind that here
 we have an expansion to the sum of orthogonal subspaces $L^U =L''_U 
\bigoplus\Proj_{L^U} L^V$ where $L''_U \neq\emptyset$. We shall define the 
operator $\Dif$ by a very similar way as in previous case:
$$
\begin{array}{l}
\Dif_{L^U >L^V} =\bigotimes\limits_j \left[\GenTimeArg_j^{-1}\Gen_j^{-1}  
(\Inv''_{j,U} \ I_{y_j} )^{t_j}
\Rest^{Z_j}_j
\right]\Change\\
\qquad {}\bigotimes\limits_j \left[\Rev^{Z_j}_j
(I_{y_j} \ \Inv''_{j,U} )^{t_j}\Gen_j \GenTimeArg_j \right]
\end{array}
$$
where the definition of $\Inv''_U$ inverting $L''_U$ looks like $\Dist_{ort}$ 
only $L''_U$ will play a role of $L'$:
$$
\Inv''_U =\Check\left[\bigotimes\limits_{k}\tilde\Res^V_k \right]\tilde\Si_{\neq\w}
\left[\bigotimes\limits_{k}\tilde\Re^V_k \right]\Check .
$$
Here $\tilde\Re^V$ and $\tilde\Res^V$ act like $\Rev^V$ and $\Rest^V$ only if
 $\a_U =1$ and if $\a_U =0$ then they do nothing, $\tilde\Si_{\neq\w}$ changes 
a sign only in one case: if $\a_U =1$ and at least $3/4$ of all frequencies 
$\w_k$ are far from $\w$: 
$|\w_k -\w |\geq 1/M$. Thus in $\Dif$ operator we shall use a set of 
ancillary registers enumerated by pairs of indices $j,k$. 

For $\bar a_j \in\Proj_{L^U} L^V$ in view of $\mu_V <\sqrt{2/3}$ an operator 
$\tilde\Si_{\neq\w}$ does not change a sign because here the fraction of all 
frequencies closed to $\w$ is $\frac{7}{8}\cdot\frac{1}{3}=\frac{7}{24}>
\frac{1}{4}$.

For $\bar a_j \bot\Proj_{L^U} L^V$ an operator $\Inv''_U$ makes nothing.

\nn

{\bf Definition of} $\Dif_{L^U >L^V}^{ort}$
\n

Suppose that $\dim L^U >\dim L^V$ and $\mu_V >\sqrt{1/3}$. The definition of 
$\Dif$ 
will be similar to the previous case only the whole subspace $L_U$ will play 
a role of $L'$:
$$
\begin{array}{l}
\Dif_{L^U >L^V}^{ort} =\bigotimes\limits_j \left[\GenTimeArg_j^{-1}\Gen_j^{-1}  
(\Inv_{j,U} \ I_{y_j} )^{t_j}
\Rest^{Z_j}_j
\right]\Change\\
\qquad {}\bigotimes\limits_j \left[\Rev^{Z_j}_j
(I_{y_j} \ \Inv_{j,U} )^{t_j}\Gen_j \GenTimeArg_j \right]
\end{array}
$$
where 
$$
\Inv_U =\Check\left[\bigotimes\limits_{k}\tilde\Res^V_k \right]\tilde\Si^{ort}_{\neq\w}
\left[\bigotimes\limits_{k}\tilde\Re^V_k \right]\Check .
$$
Here $\tilde\Si^{ort}_{\neq\w}$ changes a sign if more than a half of 
frequencies are far from $\w$: $|\w_j -\w |>1/M$. The satisfying of the 
conditions required for $\Dif$ operator is based now on inequality 
$\frac{7}{8}\frac{2}{3}=\frac{7}{12}>\frac{1}{2}$ and can be checked 
straightforwardly. 

At last estimate the complexity of constructed procedure. An operator
$\Turn$ (\ref{Turn}) requires of order $\Turn_{complexity} =M\sqrt{1/d}$ 
elementary steps. Then, $\Difference$ (\ref{Difference}) requires of 
order $\Turn_{complexity}\sqrt N$ that is $O(M\sqrt{N/d})$ elementary steps. 
Note that there exists the similar form of operator $\Difference$ which does not 
act on resulting qubit $\a_{dif}$ but changes a sign instead and such operator 
can be constructed similarly. Denote this operator by $\Difference_{sign}$. 
Assume that an input of it contains a frequency $\w$. 

\subsection{Recognition of electronic devices circuits}

Now we are ready to take up the recognition of circuits. We assume that for 
every pair of circuits for their transformations $U_1 , U_2$ subspaces spanned 
by corresponding eigenvalues are either coincident or $d$- distinguishable. 
Assume also
that our coding procedure gives one-to-one correspondence between circuits 
and $T$ basic states $e_0 , e_1 , \ldots , e_{T-1}$ in the space $H_{cir}$. 
Recognizing procedure is denoted by $\Rec$ and will have the GSA form: 

\begin{equation}
\Rec =(I_{\tilde 0}I_U )^t ,\ \ t=O(\sqrt{T})
\label{Rec}
\end{equation}
acting on states of the form $|\chi\rangle$ where basic states for $\chi$ are 
codes of circuits. Here $\tilde 0\in H_{cir}$ is chosen arbitrarily and $I_U$ 
inverts a sign of every code which circuit induces a given operator $U$. An
 implementation of
 $I_{\tilde 0}$ is straightforward and all that we need is to build $I_U$.

We define $I_U$ as
$$
I_U =\bigotimes\limits_j \left[\Conc^{-1}_{freq,\ j} 
\Difference_j  \right]\Sign\bigotimes\limits_j \left[\Difference_j  
\Conc_{freq,\ j}\right]
$$
where $\Conc_{freq}$ for every basic state $C$ of argument will generate 
some arbitrary distribution of amplitude on ancillary register with $Q$ basic
 states and then will concentrate substantial part of amplitude on a frequency 
$\w$ for which $L^U$ and $L^V$ are distinguishable (if such frequency exists).
 Then operator $\Difference_j$ changes resulting qubit for $j$th copy if and 
only if on this frequency these subspaces are distinguishable. The following 
operator $\Sign$
changes a sign if and only if at least one fifth of resulting qubits 
$\a_{dif}$ contain 1, e.g. if and 
only if operators $U$ and $U_C$ are the same.
 Then the following applications of $\Difference_j$ to each copy of register
 clean the corresponding resulting qubits and inverses operators for $\Conc_j$
 restore an initial state of ancillary register. $\Difference$ was constructed
 in the previous section and all we need now is to build $\Conc_{freq,\ j}$. 
This transformation can be defined as 

\begin{equation}
\Conc_{freq,\ j}=\GenTimeFreq_j^{-1}\ \GenFreq_j^{-1}\ (\Difference_{sign}
I_{\w_j})^{t_j}\ \GenFreq_j \ \GenTimeFreq_j
\label{Conc}
\end{equation}

If $U$ and 
$U_C$ are different then by our assumption for some $\w$ their subspaces $L^U$
 and $L^V$ are $d$-distinguishable then $\Conc_j$ will concentrate substantially
 large part of amplitude over all $j$ on some combination of such values $\w$.
Thus we have constructed a required procedure $\Rec$ which gives a target code 
with substantial probability as a result of observation of the register for 
code $C$. After the observation we can verify a fitness of a code found $C$
 by a straightforward procedure. It is similar to $I_U$ with the single change:
 $\Sign$ will be replaced by change in a special ancilla which can be observed
 after procedure and thus we shall learn does a code $C$ at hand fit or not.  

What is the complexity of our procedure $\Rec$? The complexity $Mn^2 \sqrt{N/d}$
 of $\Difference$ must be multiplied by $\sqrt{Q}$ issued from (\ref{Conc}) and 
by $\sqrt{T}$ issued from the immediate definition (\ref{Rec}). The resulting 
complexity will be $Mn^2 \sqrt{TQN/d}$. 

\subsection{Advantages of the recognizing algorithms}

Advantages of the proposed algorithms are connected with their high speed and 
small memory. Particularly, the algorithm for molecular structures recognition 
makes possible to recognize molecular circuits using microscopic memory whereas 
classically this task requires exponentially large memory.
Compare the proposed algorithms with their classical counterparts. We shall 
omit logarithmic multipliers.

1. Recognition of eigenvalues and finding thermodynamic functions. Fix some
 value of $M$ determining a precision 
of eigenvalue approximation. Consider at first
 the case when the number of ancillary qubits in a quantum gate array at hand
 is small. Then by the direct classical method we must build a matrix of 
unitary transform 
induced by a gate array. It requires of order $N^3$ steps and at least of 
order $N^2$ bits. The known quantum algorithm given by Travaglione and Milburn
 in \cite{TM} basing plainly on Abrams and Lloyd operator $\Rev$ contains 
repeated measurements of frequencies hence it requires the time of order
 $NM$ - for sparse spectrums it is of the same order as for Hams-Raedt algorithm and its only advantage over the last is exponential memory saving. 

Our algorithm recognizes an eigenvalue in $\sqrt{N} Mn$ steps. This time for the sparse area of spectrum is about
 square root of the time of best known algorithms. Here the memory will be of order 
$g^2$ qubits ($g$ is the size of gate array), that is about squared memory
 used in \cite{AL} but still exponentially smaller than of classical methods.
 Thus the proposed algorithm gives essential speedup over known methods in
 case when the number of ancillary qubits in a given gate array is small (as 
in case of molecular structure simulated by gate array) and an area of spectrum 
at hand is sparse. The same advantage 
we have with the proposed method of finding thermodynamic functions. 

If spectrums are dense we assume that $M=N$ which means that eigenvalues differ at least on $1/N$. Then the time of our algorithm is $O(N)$. 

 Consider the case when the number $a$ of ancillary qubits involved 
simultaneously in the gate array is much more than the length $n$ of input. 
Then the direct classical method requires more than $2^{2a}$ steps and at
 least $2^m$
 bits whereas our algorithm requires only about $g2^n$ steps and $gn^2$ memory 
and the quantum speedup may be more than square root. 

\n

2. The recognition of molecular structures. At first assume that spectrums are sparse. To be able to compare our method
 with the evident classical algorithm let us assume that a code of molecular 
circuit of the length $n$ is a string of ones and zeroes of this length. Thus $M=N$. The 
next natural assumption that may be also presumed for electronic circuits is 
that the sampling of a code of circuit from the uniform distribution induces a
 sampling of all possible spectrums from the uniform distribution as well. 
Then the number of all possible choices of spectrums approximations (or parts
 of spectrum subject to the statement of recognition problem) in within 
$1/L$ consisting of frequencies of the form $l/M$ is about $2^M =N$. It means 
that in our assumption $M$ and $Q$ must be logarithmic of $N$. Hence our method has the time complexity $O(N)$. With these assumptions 
the time complexity of the classical direct algorithm examining all codes and 
calculating the corresponding spectrums is about $N^3 \cdot N=N^4$ whereas our 
algorithm requires the time about $N$ and logarithmic memory. Thus the quantum 
time for this problem is about fourth root of the time of classical direct
 method and quantum space is logarithmic. 

If spectrums are dense then $Q$ and $M$ will be of order $N$ and our method requires the time $O(N^{2.5} )$ comparatively with $O(N^4 )$ of direct classical way. 

 3. Recognition of electronic devices. Here in the general case there are no 
classical analogs. 
Compare two algorithm constructed above with their classical and known 
quantum counterparts. At first consider the single recognizing quantum 
algorithm that can be easily deduced from the technique known before. This 
is an algorithm of recognizing a circuit realizing classical involutive 
function of the form $f:\ Q\ar Q,\ f=f^{-1}$. This task can be reduced to
 the search of $y$ such that the following logic formula is true: $\forall 
x\ A(x,y)$ where $A(x,y)$ is some predicate. Indeed, if we take $Y(x)=U(x)$
 in place of $A(x,y)$ where $Y$ is a function whose code is $y$ then we just obtain the problem of recognition of circuit generating $U$.
 An algorithm for such formulas given in \cite{BCW} has the time complexity of
 order $\sqrt{TN}$. This task is a particular case of our algorithm for 
involutive devices and it has the same complexity. 
In this particular case quantum time is of order square root of classical. But 
if we regard a bit more general but still restricted problem of
 recognition of involutive devices producing linear combinations of 
basic states (like quantum subroutines) an advantage over classical method of
 recognition will be more. For example, consider the restricted problem when 
we must choose between two alternative constructions of a tested device 
inducing not classical unitary transformation. The naive method of observing
 the results of action of a tested device on the different inputs requires of
 order $\frac{1}{\e} N^3$ steps to restore the matrix of the operator $U_C$ in
 within $\e$. Then this $\e$ must be less than $1/\sqrt{N}$ to give vanishing  
difference between operators in Hilbert space. Hence the time complexity of 
the naive method of recognition is roughly $N^{7/2}$. On the other hand the
 method proposed in the section 3.4 requires the choice of $d$ only converging 
to zero with $N$ converging to infinity. Thus the time required by our method is 
a little more than $\sqrt{N}$. We thus have almost seventh degree speedup for
 the problem of distinguishing electronic circuits generating transformations 
with not classical matrices. 

\section{Conclusion}

The main conclusion is that molecular structure and physical properties of environment can be 
quickly recognized on the microscopic level whereas the classical methods
 require huge time and especially memory. The new algorithms recognizing 
eigenvalues with fixed precision and molecular structure, finding
 thermodynamic functions give a quadratic speedup comparatively with 
the best classical algorithms and exponential memory saving. The new 
method based 
on quantum computing was proposed for fast recognition of electronic devices. By this method two devices with the same given 
spectrum can be distinguished in the time about seventh root of the time 
of direct measurements. All these algorithms show essential potential 
advantages of microscopic sized quantum devices comparatively with their
 classical counterparts with much bigger memory. The advantages touch
 intellectual tasks like recognition of the structure of other devices 
and important properties of environment. The proposed algorithms are built 
of standard known subroutines; they have simple structure and lay completely
 in the framework of conventional paradigm of quantum computing.  

\section{Acknowledgements}

I am sincerely grateful to Kamil Valiev for the creating of conditions for 
investigations in quantum computing in the Institute of Physics and Technology 
and for his attention and valuable advices concerning my work.

\end{document}